\documentclass[sigconf,nonacm]{acmart}

\usepackage[utf8]{inputenc}
\usepackage[english]{babel}
\usepackage{xspace}
\usepackage{enumitem}

\AtBeginDocument{%
  \providecommand\BibTeX{{%
    \normalfont B\kern-0.5em{\scshape i\kern-0.25em b}\kern-0.8em\TeX}}}

\setcopyright{acmcopyright}
\copyrightyear{2022}
\acmYear{2022}
\acmDOI{}
\acmConference[ISSTA 2022]{ACM SIGSOFT International Symposium on Software Testing and Analysis}{18-22 July, 2022}{Daejeon, South Korea}
\acmPrice{15.00}
\acmISBN{978-1-4503-XXXX-X/18/06}

\begin{document}

\title[SnapFuzz: An Efficient Fuzzing Framework for Network Applications]{SnapFuzz: An Efficient Fuzzing Framework \\ for Network Applications}

\author{Anastasios Andronidis}
\affiliation{
  \institution{Imperial College London}
  \city{London}
  \country{United Kingdom}}
\email{a.andronidis@imperial.ac.uk}

\author{Cristian Cadar}
\affiliation{
  \institution{Imperial College London}
  \city{London}
  \country{United Kingdom}}
\email{c.cadar@imperial.ac.uk}

\newcommand{\snapfuzz}{\textit{SnapFuzz}\xspace}

\newcommand{\afl}{\textit{AFL}\xspace}
\newcommand{\aflnet}{\textit{AFLNet}\xspace}
\newcommand{\libfuzzer}{\textit{LibFuzzer}\xspace}

\newcommand{\asan}{\textit{ASan}\xspace}
\newcommand{\ubsan}{\textit{UBSan}\xspace}
\newcommand{\tsan}{\textit{TSan}\xspace}

\newcommand{\lightftp}{LightFTP\xspace}
\newcommand{\dcmqrscp}{Dcmqrscp\xspace}
\newcommand{\dnsmasq}{Dnsmasq\xspace}
\newcommand{\live}{LIVE555\xspace}
\newcommand{\tinydtls}{TinyDTLS\xspace}

\newcommand{\preeny}{Preeny\xspace}
\newcommand{\multifuzz}{MultiFuzz\xspace}

\newcommand{\code}[1]{\texttt{#1}}

\newcommand{\dcmqrscpAnT}{17 hours 35 minutes\xspace}
\newcommand{\dcmqrscpAnM}{1055\xspace}

\newcommand{\dnsmasqAnT}{15 hours 17 minutes\xspace}
\newcommand{\dnsmasqAnM}{917\xspace}

\newcommand{\tinydtlsAnT}{23 hours 21 minutes\xspace}
\newcommand{\tinydtlsAnM}{1401\xspace}

\newcommand{\lightftpAnT}{35 hours 35 minutes\xspace}
\newcommand{\lightftpAnM}{2135\xspace}

\newcommand{\liveAnT}{25 hours 47 minutes\xspace}
\newcommand{\liveAnM}{1547\xspace}

\newcommand{\dcmqrscpSfT}{2 hours 7 minutes\xspace}
\newcommand{\dcmqrscpSfM}{127\xspace}

\newcommand{\dnsmasqSfT}{30 minutes\xspace}
\newcommand{\dnsmasqSfM}{30\xspace}

\newcommand{\tinydtlsSfT}{34 minutes\xspace}
\newcommand{\tinydtlsSfM}{34\xspace}

\newcommand{\lightftpSfT}{34 minutes\xspace}
\newcommand{\lightftpSfM}{34\xspace}

\newcommand{\liveSfT}{63 minutes\xspace}
\newcommand{\liveSfM}{63\xspace}

\newcommand{\dcmqrscpSfAnSu}{8.4x\xspace}
\newcommand{\dnsmasqSfAnSu}{30.6x\xspace}
\newcommand{\tinydtlsSfAnSu}{41.2x\xspace}
\newcommand{\lightftpSfAnSu}{62.8x\xspace}
\newcommand{\liveSfAnSu}{24.6x\xspace}

\newcommand{\dcmqrscpSfPAnSu}{1.30x\xspace}
\newcommand{\dnsmasqSfPAnSu}{1.90x\xspace}
\newcommand{\tinydtlsSfPAnSu}{3.40x\xspace}
\newcommand{\lightftpSfPAnSu}{1.90x\xspace}
\newcommand{\liveSfPAnSu}{3.00x\xspace}

\newcommand{\dcmqrscpSfAAnSu}{3.85x\xspace}
\newcommand{\dnsmasqSfAAnSu}{3.47x\xspace}
\newcommand{\tinydtlsSfAAnSu}{12.21x\xspace}
\newcommand{\lightftpSfAAnSu}{1.79x\xspace}
\newcommand{\liveSfAAnSu}{5.93x\xspace}

\newcommand{\dcmqrscpSfSlAnSu}{1.00x\xspace}
\newcommand{\dnsmasqSfSlAnSu}{1.00x\xspace}
\newcommand{\tinydtlsSfSlAnSu}{1.00x\xspace}
\newcommand{\lightftpSfSlAnSu}{2.76x\xspace}
\newcommand{\liveSfSlAnSu}{1.04x\xspace}

\newcommand{\dcmqrscpSfStAnSu}{1.00x\xspace}
\newcommand{\dnsmasqSfStAnSu}{1.00x\xspace}
\newcommand{\tinydtlsSfStAnSu}{1.00x\xspace}
\newcommand{\lightftpSfStAnSu}{1.00x\xspace}
\newcommand{\liveSfStAnSu}{1.04x\xspace}

\newcommand{\dcmqrscpSfDAnSu}{1.94x\xspace}
\newcommand{\dnsmasqSfDAnSu}{4.79x\xspace}
\newcommand{\tinydtlsSfDAnSu}{1.09x\xspace}
\newcommand{\lightftpSfDAnSu}{2.39x\xspace}
\newcommand{\liveSfDAnSu}{1.25x\xspace}

\newcommand{\dcmqrscpSfFsAnSu}{1.55x\xspace}
\newcommand{\dnsmasqSfFsAnSu}{1.00x\xspace}
\newcommand{\tinydtlsSfFsAnSu}{1.09x\xspace}
\newcommand{\lightftpSfFsAnSu}{2.23x\xspace}
\newcommand{\liveSfFsAnSu}{1.18x\xspace}

\newcommand{\minSuApp}{\dcmqrscp}
\newcommand{\minSu}{\dcmqrscpSfAnSu}
\newcommand{\minAnT}{\dnsmasqAnT}

\newcommand{\maxSuApp}{\lightftp}
\newcommand{\maxSu}{\lightftpSfAnSu}
\newcommand{\maxAnT}{\lightftpAnT}

\newcommand{\meanSuApp}{\live}
\newcommand{\meanSu}{\liveSfAnSu}

\newcommand{\minSfT}{\tinydtlsSfT}
\newcommand{\maxSfT}{\dcmqrscpSfT}

\newcommand{\lightftpSfAsanT}{1 hour 58 minutes\xspace}
\newcommand{\lightftpSfAsanM}{118\xspace}

\newcommand{\dcmqrscpSfAsanT}{x minute\xspace}
\newcommand{\dcmqrscpSfAsanM}{x\xspace}

\newcommand{\dnsmasqSfAsanT}{x minutes\xspace}
\newcommand{\dnsmasqSfAsanM}{x\xspace}

\newcommand{\liveSfAsanT}{x minutes\xspace}
\newcommand{\liveSfAsanM}{x\xspace}

\newcommand{\tinydtlsSfAsanT}{x minutes\xspace}
\newcommand{\tinydtlsSfAsanM}{x\xspace}

\newcommand{\lightftpSfAsanSu}{18.4x\xspace}
\newcommand{\dcmqrscpSfAsanSu}{x\xspace}
\newcommand{\dnsmasqSfAsanSu}{x\xspace}
\newcommand{\liveSfAsanSu}{x\xspace}
\newcommand{\tinydtlsSfAsanSu}{x\xspace}

\newcommand{\ie}{i.e.\ }
\newcommand{\eg}{e.g.\ }
\newcommand{\etal}{et al.}
\newcommand{\vs}{vs.\ }
\newcommand{\etc}{etc.\ }

\newcommand{\tasos}[1]{\textcolor{green}{AA: #1}}
\newcommand{\CC}[1]{\textcolor{blue}{CC: #1}}

\begin{abstract}
  In recent years, fuzz testing has benefited from increased computational power and important algorithmic advances, leading to systems that have discovered many critical bugs and vulnerabilities in production software.
  Despite these successes, not all applications can be fuzzed efficiently.
  In particular, stateful applications such as network protocol implementations are constrained by their low fuzzing throughput and the need to develop fuzzing harnesses that reset their state and isolate their side effects.

  In this paper, we present \snapfuzz, a novel fuzzing framework for network applications.
  \snapfuzz offers a robust architecture that transforms slow asynchronous network communication into fast synchronous communication, snapshots the target at the latest point at which it is safe to do so, speeds up all file operations by redirecting them to a custom in-memory filesystem, and removes the need for many fragile modifications, such as configuring time delays or writing clean-up scripts, together with several other improvements.

  Using \snapfuzz, we fuzzed five popular networking applications: \lightftp, \tinydtls, \dnsmasq, \live and \dcmqrscp.
  We report impressive performance speedups of \lightftpSfAnSu, \tinydtlsSfAnSu, \dnsmasqSfAnSu, \liveSfAnSu, and \dcmqrscpSfAnSu, respectively, with significantly simpler fuzzing harnesses in all cases.
  Through its performance advantage, \snapfuzz has also found 12 extra crashes compared to \aflnet in these applications.

\end{abstract}

\maketitle

\section{Introduction}
\label{sec:intro}
Fuzzing is an effective technique for testing software systems, with popular fuzzers such as \afl and \libfuzzer having found thousands of bugs in both open-source and commercial software.
For instance, Google has discovered over 25,000 bugs in their products and over 22,000 bugs in open-source code using greybox fuzzing~\cite{clusterfuzz-trophies}.

Unfortunately, not all software can benefit from such fuzzing campaigns.
One important class of software, network protocol implementations, are difficult to fuzz.
There are two main difficulties: the fact that in-depth testing of such applications needs to be aware of the network protocol they implement (\eg FTP, DICOM, SIP), and the fact that they have side effects, such as writing data to the file system or exchanging messages over the network.

There are two main approaches for testing such software in a meaningful way.
One approach, adopted by Google's \textit{OSS-Fuzz}, is to write unit-level test drivers that interact with the software via its API~\cite{libFuzzer}.
While such an approach can be effective, it requires significant manual effort, and does not perform system-level testing where an actual server instance interacts with actual clients.

A second approach, used by \aflnet~\cite{aflnet}, performs system-level testing by starting actual server and client processes, and generating random message
exchanges between them which nevertheless follow the underlying network protocol.
Furthermore, it does so without needing a specification of the protocol, but rather by using a corpus of real message exchanges between server and clients.
\aflnet's approach has significant advantages, requiring less manual effort and performing end-to-end testing at the protocol level.

While \aflnet makes important advances in terms of fuzzing network protocols, it has two main limitations.
First, it requires users to add or configure various time delays in order to make sure the protocol is followed, and to write clean-up scripts to reset the state across fuzzing iterations.
Second, it has poor fuzzing performance, caused by asynchronous network communication, various time delays, and expensive file system operations, among others.

\snapfuzz addresses both of these challenges thorough a robust architecture that transforms slow asynchronous network communication into fast synchronous communication, speeds up file operations and removes the need for clean-up scripts via an in-memory filesystem, and improves other aspects such as delaying and automating the  forkserver placement, correctly handling signal propagation and eliminating developer-added delays.

These improvements significantly simplify the construction of fuzzing harnesses for network applications and dramatically improve fuzzing throughput in the range of \minSu to \maxSu (mean: \meanSu) for a set of five popular server benchmarks.

\section{From AFL to AFLNet to SnapFuzz}
\label{sec:afl-aflnet-snapfuzz}
In this section, we first discuss how \afl and \aflnet work, focusing on their internal architecture and performance implications, and then provide an overview of \snapfuzz's architecture and main contributions.

\subsection{American Fuzzy Lop (\afl)}
\label{sec:afl}

\begin{figure}
  \centering
  \resizebox{0.8\columnwidth}{!}{
    \includegraphics{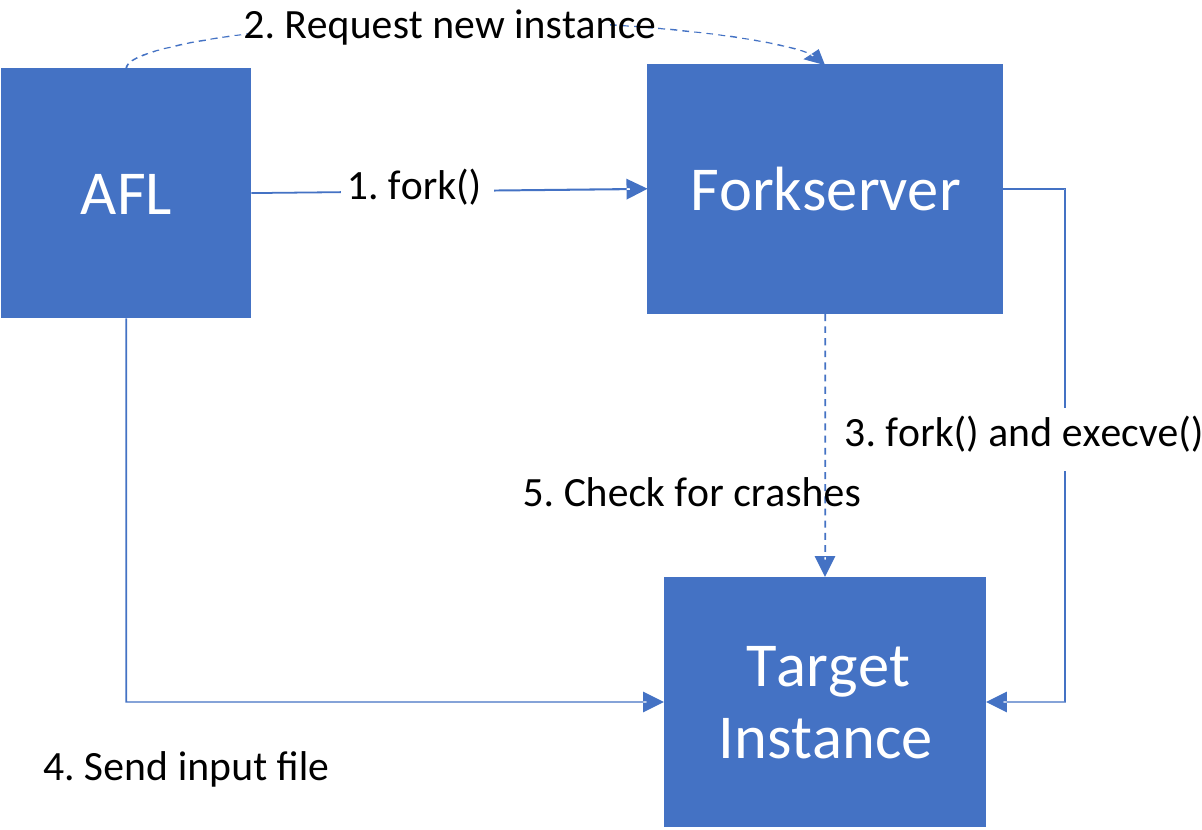}
  }
  \caption{Architecture of \afl's forkserver mode.}
  \captionsetup{justification=centering}
  \label{fig:arch-afl}
\end{figure}

\afl~\cite{afl-fuzz} is a greybox fuzzer that uses an effective coverage-guided genetic algorithm.
\afl uses a modified form of edge coverage to efficiently identify inputs that change the target application's control flow.

In a nutshell, \afl first loads user-provided initial seed inputs into a queue, picks an input, and mutates it using a variety of strategies.
If a mutated input covers a new state, it is added to the queue and the cycle is repeated.

At a systems level, \afl's simplest mode (called \textit{dumb} mode) is to restart the target application from scratch by forking first and then creating a fresh process via \code{execve}.
When this happens, the standard sequence of events to start a process is taking place, with the OS loader first initializing the target application and its libraries into memory.
\afl then sends to the new process the fuzzed input through a file descriptor that usually points to an actual file or \code{stdin}.
Lastly, \afl waits for the target to terminate, but kills it if a predefined timeout is exceeded.
These steps are repeated for every input \afl wants to provide to the target application.

\afl's dumb mode is rather slow as too much time is spent on loading and initialising the target and its libraries (such as \code{libc}) for every generated input.
Ideally, the application would be restarted after all these initialisation steps are done, as they are irrelevant to the input provided by \afl.
This is exactly what \afl's \textit{forkserver mode} offers, as shown in Figure~\ref{fig:arch-afl}.

In this mode, AFL first creates a child server called the \textit{forkserver} (step 1 in Figure~\ref{fig:arch-afl}), which loads the target application via \code{execve} and freezes it just before the \code{main} function is about to start.

Then, in each fuzzing iteration, the following steps take place in a loop: \afl requests a new target instance from the forkserver (step~2), the forkserver creates a new instance (step~3), \afl sends fuzzed input to this new instance (step~4), and the forkserver checks the target instance for crashes (step~5).


With this forkserver snapshotting mechanism, \afl replaces the loading overhead by a much less expensive \code{fork} call, while guaranteeing that the application will be at its initial state for every freshly generated input from \afl.
In the most recent versions of \afl, this is implemented as an LLVM pass, but other methods that do not require access to the source code are also available.

One additional optimisation that \afl offers is the \textit{deferred forkserver mode}.
In this mode, the user can manually add in the target's source code a special call to an internal function of \afl in order to instruct it to create the forkserver  at a later stage in the execution of the target application.
This can provide significant performance benefits in the common case where the target application needs to perform a long initialisation phase before it is able to consume \afl's input.
Unfortunately though, this mode requires the user not only to have access to the source code of the target application, but also knowledge of the internals of the target application in order to place the deferred call at the correct stage of execution.  As we will explain in \S\ref{sec:deferred-forkserver}, the forkserver placement has several restrictions (\eg it cannot be placed after file descriptors are created) and if these restrictions are violated, the fuzzing campaign can waste a lot of time exploring invalid executions.

\subsection{\aflnet}
\label{sec:aflnet}

\begin{figure}
  \centering
  \resizebox{0.8\columnwidth}{!}{
    \includegraphics{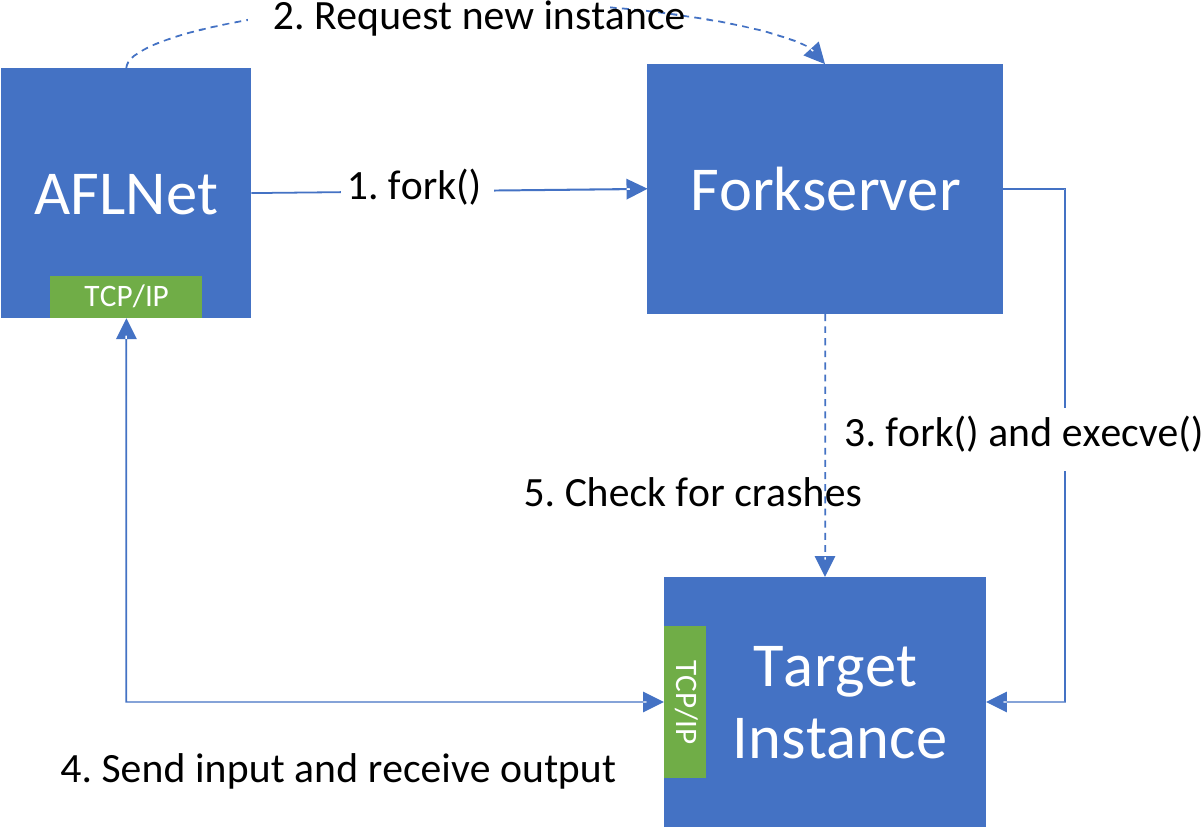}
  }
  \caption{Architecture of \aflnet.}
  \captionsetup{justification=centering}
  \label{fig:arch-aflnet}
\end{figure}

\afl essentially targets applications that receive inputs via files (with \code{stdin} a special file type).
This means that it is not directly applicable to network applications, as they expect inputs to arrive through network sockets and follow an underlying \textit{network protocol}.


\aflnet~\cite{aflnet} extends \afl to work with network applications.
Its most important contribution is that it proposes a new algorithm on how to generate inputs that follow the underlying network protocol (\eg the FTP, DNS or SIP protocols).
More specifically, \aflnet infers the underlying protocol via examples of recorded message exchanges between a client and the server.

\aflnet also extends \afl by building the required infrastructure to direct the generated inputs through a network socket to the target application, as shown in Figure~\ref{fig:arch-aflnet}.
More precisely, from a systems perspective, \aflnet acts as the client application.
After a configurable delay waiting for the server under fuzzing to initialize, it sends inputs to the server through TCP/IP or UDP/IP sockets, with configurable delays between those deliveries (we describe the various time delays needed by \aflnet in \S\ref{sec:snapfuzz-protocol}).
\aflnet consumes the replies from the server (or else the server might block) and also sends to the server a \code{SIGTERM} signal after each exchange is deemed complete, as usually network applications run in infinite loops.

As shown in Figure~\ref{fig:arch-aflnet}, the architecture of \aflnet is similar to that of \afl's deferred forkserver mode, except that communication takes place over the network instead of via files.

Network applications like databases or FTP servers are often stateful, keeping track of their state by storing information to various files.
This can create issues during a fuzzing campaign because when \aflnet restarts the application, its state might be tainted by information from a previous execution.
To avoid this problem, \aflnet requires the user to write custom \textit{clean-up scripts} that are invoked to reset any filesystem state.

We use the term \textit{fuzzing harness} to refer to all the code that users need to write in order to be able to fuzz an application.
In \aflnet, this includes the client code, the various time delays that need to be manually added, and the clean-up scripts.  One important goal of \snapfuzz is to simplify the creation of fuzzing harnesses for network applications.


\subsection{\snapfuzz}
\label{sec:snapfuzz}

\begin{figure}
  \centering
  \resizebox{0.9\columnwidth}{!}{
    \includegraphics{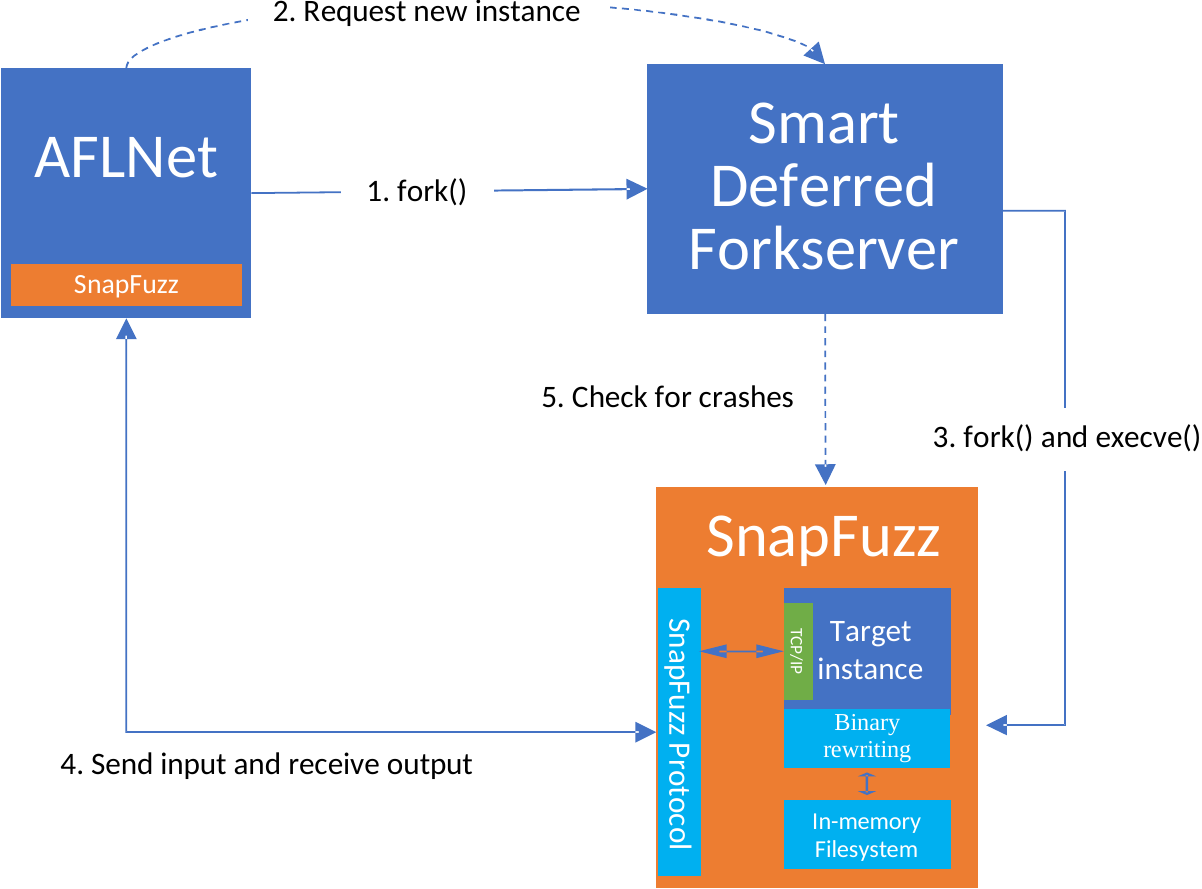}
  }
  \caption{Architecture of \snapfuzz.\vspace{-0.5cm}}
  \captionsetup{justification=centering}
  \label{fig:arch-sf}
\end{figure}

\snapfuzz is built on top of \aflnet by revamping its networking communication architecture as shown in Figure~\ref{fig:arch-sf}, without any modifications to \aflnet's fuzzing algorithm.

\snapfuzz's main goals are (1)~to improve the performance (throughput) of fuzzing network applications, and
(2)~lower the barrier for testing network applications by simplifying the construction of fuzzing harnesses, in particular by eliminating the need to add manually-specified time delays and to write clean-up scripts.
At the same time, it is not a goal of \snapfuzz to improve in any way \afl's and \aflnet's fuzzing algorithms or mutation strategies.

At a high level, \snapfuzz achieves its significant performance gains by: 
optimising all networking communications by eliminating synchronisation delays (\textit{the \snapfuzz protocol)};
automatically injecting \afl's forkserver deeper into the application than otherwise possible and without the user's intervention (\textit{smart deferred forkserver});
performing binary rewriting-enabled optimisations which eliminate additional delays and inefficiencies;
automatically resetting any filesystem state;
and optimising filesystem writes by redirecting them into an in-memory filesystem.

\snapfuzz also makes fuzzing harness development easier and in some cases trivial by completely removing the need for manual code modifications.
Such manual changes are often required to: reset the state of either the target or its environment after each fuzzing iteration; terminate the target, as usually servers run in infinite loops; pin the CPU for threads and processes; and add deferred forkserver support to the target.



Figure~\ref{fig:arch-sf} shows the architecture of \snapfuzz.
While at a high-level it resembles that of \aflnet, there are several important changes.
First, \snapfuzz intercepts the external actions of the target application using \textit{binary rewriting} (\S\ref{sec:binary-rewriting}).
It then monitors the behaviour of both the target application and the \aflnet client in order to eliminate synchronisation delays using its \snapfuzz protocol (\S\ref{sec:snapfuzz-protocol}).
Second, a custom in-memory filesystem is added, to improve performance and facilitate resetting the state after each fuzzing iteration (\S\ref{sec:state-reset}).
Third, the forkserver is replaced by a smart deferred forkserver, which automates and optimizes the forkserver placement  (\S\ref{sec:deferred-forkserver}).
We describe the main components of \snapfuzz in detail in the next section.

\section{Design}
\label{sec:design}
\snapfuzz has two main goals: significantly increase fuzzing throughput, and simplify the  construction of fuzzing harnesses.

At a high-level, \snapfuzz accomplishes these goals by intercepting all the communication between the target application and its environment via binary rewriting (\S\ref{sec:binary-rewriting}).
By controlling this communication, \snapfuzz can then:

\begin{enumerate}[leftmargin=*]
\item Implement an efficient network fuzzing protocol which notifies the fuzzer when the target application is ready to accept a new request or when a response is ready to be consumed (\S\ref{sec:snapfuzz-protocol}).
      This improves fuzzing throughput and eliminates the need for all the custom delays that \aflnet users need to insert in order to synchronise the communication between the fuzzer and the target application.
            \snapfuzz also replaces internet sockets by UNIX domain sockets, which improves performance, and implements an efficient server termination strategy.

 \item Redirect all file operations to use an in-memory filesystem (\S\ref{sec:state-reset}).
   This improves the performance of filesystem operations, and obviates the need for user-provided clean-up scripts, as \snapfuzz can automatically clean up after each fuzzing iteration by simply discarding the in-memory state.

\item Automatically place and defer the forkserver (``smart deferred forkserver'') to the latest safe point (\S\ref{sec:deferred-forkserver}).
  This improves performance and eliminates the need for manual annotations.
   
\item Eliminate custom delays, unnecessary system calls and potentially expensive clean-up routines that are part of the target application, correctly propagate signals from child processes, and better control CPU affinity (\S\ref{sec:extras}).

\end{enumerate}

\subsection{Binary Rewriting}
\label{sec:binary-rewriting}

\snapfuzz implements a load-time binary rewriting subsystem that dynamically
intercepts both the OS loader's and the target's functionalities in order to monitor and modify all external behaviours of the target application.

Applications interact with the external world via \emph{system calls}, such as \code{read()} and \code{write()} in Linux, which provide various OS services.
As an optimization, Linux provides some services via \emph{vDSO (virtual Dynamic Shared Object)} calls.  vDSO is essentially a small shared library injected by the kernel in every application in order to provide fast access to some services.  For instance, \emph{gettimeofday()} is typically using a vDSO call on Linux. 

The main goal of the binary rewriting component of \snapfuzz is to intercept all the system calls and vDSO calls issued by the application being fuzzed, and redirect them to a \emph{system call handler}.  \S\ref{sec:binary-rewriting-impl} presents the implementation details. 

By intercepting the target application's interactions with its outside environment at this level of granularity, \snapfuzz can significantly increase fuzzing throughput and eliminate the need for custom delays and scripts, as we discuss in the next subsections.


\subsection{\snapfuzz Network Fuzzing Protocol: Eliminating Communication Delays}
\label{sec:snapfuzz-protocol}

\begin{figure}
      \centering
      \resizebox{1.00\columnwidth}{!}{
            \includegraphics[trim=0 0 0 20, clip]{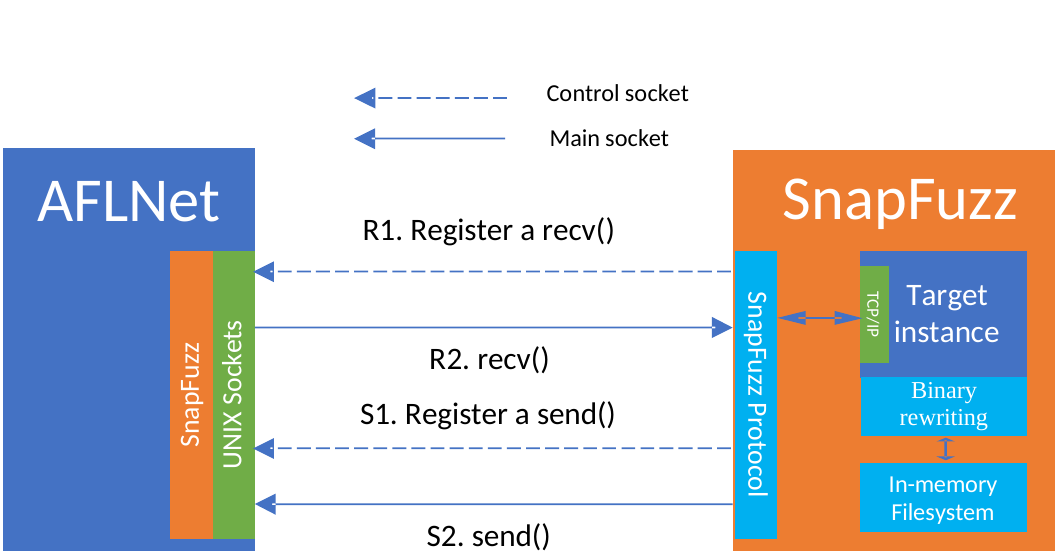}
      }
      \caption{Messages exchanged for each \texttt{recv} and \texttt{send}.}
      \captionsetup{justification=centering}
      \label{fig:sf-prot}
\end{figure}

Network applications often implement multistep protocols with multiple requests and replies per session.
One of \aflnet's main contributions is to infer the network protocol starting from a set of recorded message exchanges.
However, \aflnet cannot guarantee that during a certain fuzzing iteration the target will indeed respect the protocol.
Deviations might be possible for instance due to a partly-incorrect protocol being inferred, bugs in the target application, or most commonly due to the target not being ready to send or receive a certain message.

Therefore, \aflnet performs several checks and adds several user-specified delays  to ensure communication is in sync with the protocol.
These communication delays, which can significantly degrade the fuzzing throughput, are:

\begin{enumerate}[leftmargin=*]
      \item A delay to allow the server to initialise before \aflnet attempts to communicate.
      \item A delay specifying how long to wait before concluding that no responses are forthcoming and instead try to send more information, and
      \item A delay specifying how long to wait after each packet is sent or received.
\end{enumerate}

These delays are necessary, as otherwise the OS kernel will reject packets that come too fast while the target is not ready, and \aflnet will desynchronize from its state machine.
But they cause a lot of time to be wasted, essentially because \aflnet does not know whether the target is ready to send or receive information.

\snapfuzz overcomes this challenge through a simple but effective network fuzzing protocol.
The protocol keeps track of the next action of the target, and notifies \aflnet about it.
Figure~\ref{fig:sf-prot} shows the messages exchanged between \snapfuzz and \aflnet on each \code{recv} (for receiving data) and \code{send} (for sending data)  system calls.
Essentially, to avoid the need for the communication delays discussed above, \snapfuzz informs \aflnet when the target is about to issue a \code{recv} or a \code{send}. 
This is performed by introducing an additional \textit{control socket} (implemented via an efficient UNIX domain socket), which is used as a send-only channel from the \snapfuzz plugin to \aflnet.


The \snapfuzz network fuzzing protocol additionally implements the following two optimisations:


\medskip \noindent
\textbf{UNIX Domain Sockets.} The standard Internet sockets (TPC/IP and UDP/IP) used by \aflnet to communicate to the target and send it fuzzed inputs are unnecessarily slow.
  As observed before~\cite{multifuzz}, replacing them with UNIX domain sockets can lead to significant performance speed-ups.
  We discuss how this is achieved in \S\ref{sec:domain-sockets}.

\medskip \noindent
\textbf{Efficient Server Termination.} 
Network servers usually run in a loop.
This loop is terminated either via a special protocol-specific keyword or an OS signal.
Since \aflnet cannot guarantee that each fuzzing iteration will finish via a termination keyword, if the target does not terminate, it sends it a \code{SIGTERM} signal and waits for it to terminate.
Signal delivery is slow and also servers might take a long time to properly terminate execution.
In the context of fuzzing, proper termination is not so important, while fuzzing throughput is.
\snapfuzz implements a simple mechanism to terminate the server: when it receives an empty string, it infers that the fuzzer has no more inputs to provide and the application is instantly killed.
This obviously has the downside that it could miss bugs in the termination routines, but these could be tested separately.

In summary, the \snapfuzz network fuzzing protocol improves fuzzing performance (significantly, as shown in the evaluation) and simplifies fuzzing harness construction by eliminating the need to manually specify three different communication delays.

\subsection{Efficient State Reset}
\label{sec:state-reset}

%

\aflnet users typically have to write a clean-up script to reset the application state after each fuzzing iteration.
For instance, \lightftp under \aflnet requires a script that cleans up any directories or files that have been created in the previous iteration.
Under \snapfuzz, there is no need for such a clean-up script, which simplifies the test harness construction, and improves performance by avoiding the invocation of the clean-up script.

\snapfuzz solves this challenge by employing an in-memory filesystem.
Using the in-memory filesystem \code{tmpfs} under UNIX is a well-known optimisation in the context of fuzzing.\footnote{\url{https://www.cipherdyne.org/blog/2014/12/ram-disks-and-saving-your-ssd-from-afl-fuzzing.html}}\textsuperscript{,}\footnote{\url{https://medium.com/@dhiraj_mishra/fuzzing-vim-53d7cf9b5561}}\textsuperscript{,}\footnote{\url{https://www.cis.upenn.edu/~sga001/classes/cis331f19/hws/hw1.pdf}} 

\snapfuzz uses an in-memory filesystem both for efficiency and for removing the need for clean-up scripts involving filesystem state.
However, we are not using \code{tmpfs}, but a custom in-memory filesystem that uses the \code{memfd\_create} system call for files and the \textit{Libsqlfs} library for directories (see \S\ref{sec:in-mem-fs} for details).
This allows us to quickly duplicate state after forking, as explained below. 

In the simplest case where \afl checkpoints the target application before \code{main}, no filesystem modifications have happened at the point where the forkserver is placed.
So when a fuzzing iteration has finished, the target application process just exits and the OS discards its memory, which includes any in-memory filesystem modifications made during fuzzing.
Then, when the forkserver spawns a new instance of the target application, the filesystem is brought back to a state where all initial files of the actual filesystem are unmodified.

The situation is more complicated when the deferred forkserver is placed after the target application has already created some files.
In our implementation, which is based \code{memfd\_create}, when the forkserver creates a new instance to be fuzzed, the Linux kernel shares the memory pages associated with the newly-created in-memory files between the new instance and the forkserver.
Note that using \code{tmpfs} would not solve this issue---as far as we know, there is no way to duplicate a \code{tmpfs} filesystem in a copy-on-write way.
%
%
This sharing of pages between the new instance and the forkserver is problematic, as now any modifications to the in-memory files by the fuzzed application instance will persist even after the instance finishes execution.
So in the next iteration, when the forkserver creates a new instance, this new instance will inherit those modifications too.

\snapfuzz solves this issue as follows. 
First, note that \snapfuzz knows whether the application is executing before or after the forkserver's checkpoint, as it intercepts all system calls, including \code{fork}.
While the target application executes before the forkserver's checkpoint, \snapfuzz allows all file interactions to be handled normally. 
When a new instance is requested from the forkserver, \snapfuzz recreates in the new instance all in-memory files registered in the in-memory filesystem and copies all their contents by using the efficient \code{sendfile} system call once per in-memory file.

\subsection{Smart Deferred Forkserver}
\label{sec:deferred-forkserver}

As discussed in \S\ref{sec:afl}, the deferred forkserver can offer great performance benefits by avoiding initialisation overheads in the target.
Such overheads include loading the shared libraries used by the target, parsing configuration files and cryptographic initialisation routines.
Unfortunately, for the deferred forkserver to be used, the user needs to manually modify to source code of the target.
Furthermore, the deferred forkserver cannot be used after the target has created threads, child processes, temporary files, network sockets, offset-sensitive file descriptors, or shared-state resources, so the user has to carefully decide where to place it: do it too early and optimisation opportunities are missed, do it too late and correctness is affected.


\snapfuzz makes two important improvements to the deferred forkserver: first, it makes it possible to defer it much further than usually possible with \afl's architecture, and second, it does so automatically, without any need for manual source modifications.

The two components which enable \snapfuzz to place the forkserver after many system calls which normally would have caused problems are:
(1)~its custom network fuzzing protocol which allows it to skip network setup calls such as \code{socket} and \code{accept} (\S\ref{sec:snapfuzz-protocol}) and
(2)~its in-memory filesystem, which transforms filesystem operations into in-memory changes (\S\ref{sec:state-reset}).

Via binary rewriting, \snapfuzz intercepts each system call, and places the forkserver just before it encounters either a system call that spawns new threads (\code{clone}, \code{fork}), or one used to receive input from a client.
The reason \snapfuzz still has to stop before the application spawns new threads is that the forkserver relies on \code{fork} to spawn new instances to be fuzzed, and \code{fork} cannot reconstruct existing threads---in Linux, forking a multi-threaded application creates a process with a single thread~\cite{fork}.
As a possible mitigation, we tried to combine \snapfuzz and the \textit{pthsem / GNU pth} library~\cite{pthsem}---a green threading library that provides non-preemptive priority-based scheduling, with the green threads executing inside an event-driven framework---but the performance overhead was too high.

In particular, we used \textit{pthsem} with \lightftp, as this application has to execute two \code{clone} system calls before it accepts input.
With \textit{pthsem} support, \snapfuzz's forkserver can skip these two \code{clone} calls, as well as 37 additional system calls, as now \snapfuzz can place the forkserver just before \lightftp is ready to accept input.
However, despite this gain, the overall performance was 10\% lower than in the version of \snapfuzz without \textit{pthsem}, due to the overhead of this library.
Ideally, \snapfuzz should implement a lightweight thread reconstruction mechanism to recreate all dead threads, but this is left as future work.

\subsection{Additional Binary Rewriting-enabled Optimizations}
\label{sec:extras}

In this section, we discuss several additional optimizations performed by \snapfuzz, which are enabled by its binary rewriting-based architecture. 
They concern developer-added delays, writes to \code{stdout/stderr}, signal propagation, and CPU affinity, and highlight the versatility of \snapfuzz's approach in addressing a variety of challenges and inefficiencies when fuzzing network applications.

\subsubsection{Eliminating developer-added delays}
\label{sec:eliminate-delays}

Occasionally, network applications add sleeps or timeouts in order to avoid high CPU utilisation when they poll for new connections or data.
\snapfuzz removes these delays via binary rewriting, making those calls use a more aggressive polling model.

We also noticed that in some cases application developers deliberately choose to add sleeps in order to wait for various events.
For example, \lightftp adds a one second sleep in order to wait for all its threads to terminate.
This might be fine in a production environment, but during a fuzzing campaign such a delay is unnecessary and expensive.
\snapfuzz completely skips such sleeps by intercepting and then not issuing this family of system calls at all.

\subsubsection{Avoiding \code{stdout/stderr} writes}

By default, \afl redirects \code{stdout} and \code{stderr} to \code{/dev/null}.
This is much more performant than actually writing to a file or any other medium, as the kernel optimizes those operations aggressively.
\snapfuzz goes one step further and saves additional time by completely skipping any system call that targets \code{stdout} or \code{stderr}. 

\subsubsection{Signal Propagation}
\label{sec:signals}

Some applications use a multi-process rather than a multi-threaded concurrency model.
In this case, if a subprocess crashes with a segfault, the signal might not be propagated properly to the forkserver and the crash missed.
We stumbled upon this case with the \dcmqrscp server (\S\ref{sec:dcmqrscp}) where a valid new bug was manifesting, but \aflnet was unable to detect the issue as the main process of \dcmqrscp never checked the exit status of its child processes.

As \snapfuzz has full control of the system calls of the target, whenever a process is about to exit, it checks the exit status of its child processes too.
If an error is detected, it is raised to the forkserver.

\subsubsection{Smart affinity}
\label{sec:pinning}

\afl assumes that its targets are single-threaded and thus tries to pin the fuzzer and the target to two free CPUs.
Unfortunately, there is no mechanism to handle multi-threaded applications, other than just turning off \afl's pinning mechanism.
\snapfuzz can detect when a new thread or process is about to be spawned as both \code{clone} and \code{fork} system calls are intercepted.
This creates the opportunity for \snapfuzz to take control of thread scheduling by pinning threads and processes to available CPUs.
\snapfuzz implements a very simple algorithm that pins every newly created thread or process to the next available CPU.

\section{Implementation}
\label{sec:impl}
\snapfuzz is implemented on top of \aflnet, and targets the Linux platform.
However, the ideas in \snapfuzz could be implemented using other fuzzers and operating systems. 
Below, we provide implementation details related to binary rewriting (\S\ref{sec:binary-rewriting}), our in-memory filesystem (\S\ref{sec:in-mem-fs}), and the use of UNIX domain sockets (\S\ref{sec:domain-sockets}).

\subsection{Binary Rewriting}
\label{sec:binary-rewriting-impl}

Binary rewriting in \snapfuzz employs two major components:
1)~the rewriter module, which scans the code for specific functions, vDSO and system call assembly opcodes, and redirects them to the plugin module, and
2)~the plugin module where \snapfuzz resides.

\medskip
\noindent\textbf{Rewriter.}
\label{subsec:rewriter}
\snapfuzz is an ordinary dynamically linked executable that is provided with a path to a target application together with the arguments to invoke it with.
When \snapfuzz is launched, the expected sequence of events of a standard Linux operating system are taking place, with the first step being the dynamic loader that loads \snapfuzz and its dependencies in memory.


When \snapfuzz starts executing, it inspects the target's ELF binary to obtain information about its interpreter, which in our implementation is always the standard Linux \textit{ld} loader.
\snapfuzz then scans the loader code for system call assembly opcodes and some special functions in order to instruct the loader to load the \snapfuzz plugin.
In particular, the rewriter:
(1)~intercepts the dynamic scanning of the loader in order to append the \snapfuzz plugin shared object as a dependency, and
(2)~intercepts the initialisation order of the shared libraries in order to prepend the \snapfuzz plugin initialisation code (in the \textit{.preinit\_array}).


After the \snapfuzz rewriter finishes rewriting the loader, execution is passed to the rewritten loader in order to load the target application and its library dependencies.
As the normal execution of the loader progresses, \snapfuzz intercepts its \code{mmap} system calls used to load libraries into memory, and scans these libraries in order to recursively rewrite their system calls and redirect them to the \snapfuzz plugin.
The \snapfuzz rewriter is based on the open-source load-time binary rewriter SaBRe~\cite{sabre}.

\medskip
\noindent\textbf{Plugin.}
%
%
After the loader completes, execution is passed to the target application, which will start by executing \snapfuzz's initialisation function.
Per the ELF specification, execution starts from the function pointers of \textit{.preinit\_array}.
This is a common ELF feature used by LLVM sanitizers to initialise various internal data structures early, such as the shadow memory~\cite{ASan,MSan}.
\snapfuzz is using the same mechanism to initialise its subsystems like its in-memory filesystem before the execution starts.

After the initialisation phase of the plugin, control is passed back to the target and normal execution resumed.
At this stage, the \snapfuzz plugin is only executed when the target is about to issue a system call or a vDSO call.
When this happens, the plugin checks if the call should be intercepted, and if so, it redirects it to the appropriate handler, and then returns back control to the target.




\subsection{In-memory Filesystem}
\label{sec:in-mem-fs}

As discussed in \S\ref{sec:state-reset}, \snapfuzz redirects all file operations to use a custom in-memory filesystem.
This reduces the overhead of reading and writing from a storage medium, and eliminates the need for manually-written clean-up scripts. 


\snapfuzz implements a lightweight in-memory filesystem, which uses two distinct mechanisms, one for files and the other for directories.
For files, \snapfuzz's in-memory filesystem uses the recent \break \code{memfd\_create()} system call, introduced in Linux in 2015~\cite{memfd_create}.
This system call creates an anonymous file and returns a file descriptor that refers to it.
The file behaves like a regular file, but lives in memory.
Under this scheme, \snapfuzz only needs to specially handle system calls that initiate interactions with a file through a pathname (like the \code{open} and \code{mmap} system calls).
All other system calls that handle file descriptors are compatible by default with the file descriptors returned by \code{memfd\_create}.

When a target application opens a file, the default behavior of \snapfuzz is to
check if this file is a regular file (e.g. device files are ignored), and if
so, create an in-memory file descriptor and copy the whole contents
of the file in the memory address space of the target. 
\snapfuzz keeps track of pathnames in order to avoid reloading the same file twice. This is not only a performance optimization but also a correctness requirement, as the application might have changed the contents of the file in memory.



For directories, \snapfuzz employs the \textit{Libsqlfs} library~\cite{libsqlfs}, which implements a POSIX-style file system on top of the SQLite database and allows applications to have access to a full read/write filesystem with its own file and directory hierarchy.
\textit{Libsqlfs} simplifies the emulation of a real filesystem with directories and permissions.
\snapfuzz uses \textit{Libsqlfs} for directories only, as we observed better performance for files via \textit{memfd\_create}.

\subsection{UNIX Domain Sockets}
\label{sec:domain-sockets}

\aflnet uses the standard Internet sockets (TPC/IP and UDP/IP) to communicate to the target and send it fuzzed inputs.
The Internet socket stack includes functionality---such as calculating checksums of packets, inserting headers, routing---which is unnecessary when fuzzing applications on a single machine.



To eliminate this overhead, similarly to prior work~\cite{multifuzz}, \snapfuzz replaces Internet sockets with UNIX domain sockets.
More specifically, \snapfuzz uses Sequenced Packets sockets (\code{SOCK\_SEQPACKET}).
This configuration offers performance benefits and also simplifies the implementation.
Sequenced Packets are quite similar to TCP, providing a sequenced, reliable, two-way connection-based data transmission path for datagrams.
The difference is that Sequenced Packets require the consumer (in our case the \snapfuzz plugin running inside the target application) to read an entire packet with each input system call.
This atomicity of network communications simplifies corner cases where the target application might read only parts of the fuzzer's input due to scheduling or other delays.
By contrast, \aflnet handles this issue by exposing manually defined knobs for introducing delays between network communications.

Our modified version of \aflnet creates a socketpair of UNIX domain sockets with the Sequenced Packets type, and passes one end to the forkserver, which later passes it to the \snapfuzz plugin.
The \snapfuzz plugin initiates a handshake with the modified \aflnet, after which \aflnet is ready to submit generated inputs to the target or consume responses.

Translating network communication from Internet sockets to UNIX domain sockets is not trivial, as \snapfuzz needs to support the two main IP families of TCP and UDP which have a slightly different approach to how network communication is established.
In addition, \snapfuzz also needs to support different types of synchronous and asynchronous communication such as \code{(e)poll} and \code{select}.

For the TCP family, the \code{socket} system call creates a TCP/IP socket and returns a file descriptor which is then passed to \code{bind}, \code{listen} and finally to \code{accept}, before the system is ready to send or receive any data.
\snapfuzz monitors this sequence of events on the target and when the \code{accept} system call is detected, it returns the UNIX domain socket file descriptor from the forkserver.
\snapfuzz doesn't interfere with the \code{socket} system call and intentionally allows its normal execution in order to avoid complications with target applications that perform advanced configurations on the base socket.
This strategy is similar to the one used by the in-memory file system via the \code{memfd\_create} system call (\S\ref{sec:in-mem-fs}) in order to provide compatibility by default.

The UDP family is handled in a similar way, with the only difference that instead of monitoring for an \code{accept} system call to return the UNIX domain socket of the forkserver, \snapfuzz is monitoring for a \code{bind} system call.



\section{Evaluation}
\label{sec:eval}
We demonstrate the benefits of \snapfuzz using five popular servers that were previously used in evaluating \aflnet~\cite{aflnet}: \lightftp (\S\ref{sec:lightftp}), \dcmqrscp (\S\ref{sec:dcmqrscp}), \live (\S\ref{sec:live555}) and \tinydtls (\S\ref{sec:tinydtls}).
Our experiments show that \snapfuzz significantly improves fuzzing throughput, while at the same time reducing  the effort needed to create fuzzing harnesses.
As a result of its significant performance benefit, \snapfuzz also found 12 extra crashes compared to \aflnet in these applications.

\subsection{Methodology}
\label{sec:methodology}

Since \snapfuzz's contribution is in increasing the fuzzing throughput, our main comparison metric is the number of fuzzing iterations per second.
Note that each fuzzing iteration may include multiple message exchanges between the fuzzer and the target.
A fuzzing campaign consists of a given number of fuzzing iterations.

During a fuzzing campaign, the fuzzer's speed may vary across iterations, sometimes significantly, due to different code executed by the target.
To ensure a meaningful comparison between \snapfuzz and \aflnet, rather than fixing a time budget and counting the number of iterations performed by each, we instead fix the number of iterations and measure the execution time of each system.
We monitored standard fuzzing metrics including bug count, coverage, stability, path and cycles completed, to make sure that the \snapfuzz and \aflnet campaigns have the same (or very similar) behaviour.

We chose to run each target for one million iterations to simulate realistic \aflnet fuzzing campaigns (ranging from approximately 16 to 36 hours).
We repeated the execution of each campaign 10 times. 

For bug finding, we left \snapfuzz to run for 24 hours, three times for each benchmark.
We then accumulated all discovered crashes in a single repository.
To uniquely categorise the crashes found, we recompiled all benchmarks under \asan and \ubsan, and then grouped the crashing inputs based on the reports from the sanitizers.


\subsection{Experimental Setup}

All of our experiments were conducted on a 3.0 GHz AMD EPYC 7302P 16-Core CPU and 128 GB RAM running 64-bit Ubuntu 18.04 LTS (kernel version 4.15.0-162) with an SSD disk.
Note that using a slower HDD instead of an SDD disk would likely lead to larger gains for \snapfuzz's in-memory filesystem component.

\snapfuzz is built on top of \aflnet revision \code{0f51f9e} from January 2021 and SaBRe revision \code{7a94f83}.
The servers tested and their workloads were taken from the \aflnet paper and repository at the revision mentioned above.

We used the default configurations proposed by \aflnet for all benchmarks, with a couple of exceptions.
For the \dcmqrscp server, two changes were required: 1) we had to include a Bash clean-up script to reset the state of a data directory of the server, and 2) we had to add a wait time between requests of 5 milliseconds as we observed \aflnet to desynchronise from its target.
These changes further emphasise the fact that the clean-up scripts and delays that users need to specify when building a fuzzing harness are fragile and may need adjustment when using different machines, thus \snapfuzz's ability to eliminate their need is important.

In \tinydtls we decided to decrease the inter-request wait time from 30 to 2 milliseconds, as we noticed the \aflnet performance  was seriously suffering due to this large delay.
Again, this shows that choosing the right values for these time delays is difficult.

\subsection{Summary of Results}

\begin{table}[t]
      \centering
      \caption{Time (in minutes) to complete one million fuzzing iterations in AFLNet vs Snapfuzz.}
      \captionsetup{justification=centering}
      \label{tbl:summary}
      \begin{tabular}{l|r|r|r}
                      & \aflnet      & \snapfuzz    & Speedup         \\ \hline
            \dcmqrscp & \dcmqrscpAnM & \dcmqrscpSfM & \dcmqrscpSfAnSu \\ \hline
            \dnsmasq  & \dnsmasqAnM  & \dnsmasqSfM  & \dnsmasqSfAnSu  \\ \hline
            \tinydtls & \tinydtlsAnM & \tinydtlsSfM & \tinydtlsSfAnSu \\ \hline
            \lightftp & \lightftpAnM & \lightftpSfM & \lightftpSfAnSu \\ \hline
            \live     & \liveAnM     & \liveSfM     & \liveSfAnSu     \\
      \end{tabular}
\end{table}

Table~\ref{tbl:summary} shows a summary of the results.
In particular, it compares the average time needed by \aflnet and by \snapfuzz to complete one million iterations.
As can be seen, \aflnet takes between \minAnT to \maxAnT to complete these iterations, with \snapfuzz taking only a fraction of that time, between \minSfT and \maxSfT.
The speedups are impressive in each case, varying between \minSu for \minSuApp and \maxSu for \maxSuApp.
In all cases, we observed identical coverage statistics, bug counts, and stability numbers.

\subsection{\lightftp}
\label{sec:lightftp}

\lightftp~\cite{lightftp} is a small server for file transfers that implements the FTP protocol.
The fuzzing harness instructs \lightftp to log in a specific user, list the contents of the home directory on the FTP server, create directories, and execute various other commands for system information.

\lightftp exercises a large set of \snapfuzz's subsystems.
First, it heavily utilises the filesystem, as the probability to create directories is quite high on every iteration.
Second, it has verbose logging and writing to \code{stdout}.
Third, it has a long initialisation phase, because it parses a configuration file and then undergoes a heavyweight process of initialising x509 certificates.
And lastly, \lightftp is a multi-threaded application and has a hardcoded sleep to make sure that all of its threads have terminated gracefully.

\snapfuzz optimises all the above functionalities.
All directory interactions are translated into in-memory operations, thus avoiding context switches and device (hard drive) overheads.
\snapfuzz cancels \code{stdout} and \code{stderr} writes.
\snapfuzz's smart deferred forkserver snapshots the \lightftp server after its initialisation phase and thus fuzzing under \snapfuzz pays the initialisation overhead only once.
And lastly, \snapfuzz cancels any calls to \code{sleep} and similar system calls.

Note that \snapfuzz can place the forkserver later than it could be placed manually.
For the deferred forkserver to work properly, recall that no file descriptor must be open before the forkserver snapshots the target.
This is because the underlying resource of a file descriptor is retained after a fork happens.
This limits the area where the deferred forkserver can be placed manually.
\snapfuzz overcomes this challenge with its in-memory file system as described in \S\ref{sec:in-mem-fs} and thus it is able to place the forkserver after the whole initialisation process has finished.

The one million iterations run for \lightftp take on average \lightftpAnT under \aflnet, while only \lightftpSfT under \snapfuzz, providing a \lightftpSfAnSu speedup.


\subsection{\dcmqrscp}
\label{sec:dcmqrscp}

\dcmqrscp~\cite{dcmqrscp} is a DICOM image archive server that manages a number of storage areas and allows images to be stored and queried.
The fuzzing harness instructs the DICOM server to echo connection information back to the client, and to store, find and retrieve specific images into and from its database.

\dcmqrscp heavily exercises \snapfuzz's in-memory filesystem as on every iteration the probability to read or create files is high.
\dcmqrscp also benefits from the smart deferred forkserver, as it has a long initialisation phase in which the server  dynamically loads the \textit{libnss} library and also parses multiple configuration files that dictate the syntax and capabilities of the DICOM language.



Our signal propagation subsystem (\S\ref{sec:signals}) was able to expose a bug in \dcmqrscp which was also triggered by \aflnet but was missed because signals were not properly propagated.

The one million \dcmqrscp iterations take on average \dcmqrscpAnT to execute under \aflnet, while only \dcmqrscpSfT under \snapfuzz, providing a \dcmqrscpSfAnSu speedup.

\subsection{\dnsmasq}
\label{sec:dnsmasq}

\dnsmasq~\cite{dnsmasq} is a single-threaded DNS proxy and DHCP server designed to have a small footprint and be suitable for resource-constrained routers and firewalls.
The fuzzing harness instructs \dnsmasq to query various bogus domain names from its configuration file and then report results back to its client.

\dnsmasq is an in-memory database with very little interaction with the filesystem.
Therefore, it mainly benefits from the \snapfuzz protocol and its additional optimizations of \S\ref{sec:extras}.
Furthermore, it highly benefits from the smart deferred forkserver, as it has a long initialisation process which uses \textit{dlopen()} and performs various network-related configurations.
\dnsmasq requires approximately 1,200 system calls before the process is ready to accept input.

As for other benchmarks, a manually-placed forkserver under \aflnet could not snapshot the application at the same depth as \snapfuzz's smart deferred forkserver.
This is because \dnsmasq needs to execute a sequence of system calls to establish a network connection with \aflnet.
This sequence includes creating a socket, binding its file descriptor, calling \code{listen}, executing a \code{select} to check for incoming connections, and finally accepting the connection.
Therefore, under \aflnet, the latest possible placement of the forkserver would be just before this sequence.
Under \snapfuzz, network communications are translated into UNIX domain socket communications that don't require any of the above, and thus the smart deferred forkserver can snapshot the target right before reading the input from the fuzzer, saving a lot of initialisation time.

The one million \dnsmasq iterations take on average \dnsmasqAnT under \aflnet, while only \dnsmasqSfT under \snapfuzz, providing a \dnsmasqSfAnSu speedup.

\subsection{\live}
\label{sec:live555}

\live~\cite{live555} is a single-threaded multimedia streaming server that uses open standard protocols like RTP/RTCP, RTSP and SIP.
The fuzzing harness instructs the \live server to accept requests to serve the content of a specific file in a streaming fashion, and the server replies to these requests with information and the actual streaming data.

\live only reads files and thus no state reset script is required.
It has a relatively slim initialisation phase with the main overhead coming from the many writes to \code{stdout} with welcoming messages to users.
\live mainly benefits from the \snapfuzz protocol and the elimination of \code{stdout} writes.

\live reads its files only after the forkserver performs its snapshot.
As a result, those files are not kept in the in-memory filesystem of \snapfuzz, and are read from the actual filesystem in each iteration.
We leave as future work the optimisation of predefining a set of files to be loaded in the in-memory file system when the smart deferred forkserver kicks in, so the target could read these files from memory rather the actual filesystem.

The one million \live iterations take on average \liveAnT under \aflnet, while only \liveSfT under \snapfuzz, providing a \liveSfAnSu speedup.

\subsection{\tinydtls}
\label{sec:tinydtls}

\tinydtls~\cite{tinydtls} is a DTLS 1.2 single-threaded UDP server targetting IoT devices.
In the fuzzing harness, \tinydtls accepts a new connection and then the DTLS handshake is initiated in order for communication to be established.


The protocol followed by \aflnet has several steps, and progress to the next step is accomplished either by a successful network action or after a timeout has expired.
\tinydtls supports two cipher suites, one Eliptic Curve (EC)-based, the other Pre-Shared Keys (PSK)-based.
EC-based encryption is slow, requiring the use of a large timeout between requests, which slows down fuzzing with \aflnet considerably.
%
In addition, \aflnet includes some hardcoded delays between network interactions so that it doesn't overwhelm the target---without these delays, network packets might be dropped and \aflnet's state machine desynchronized.
Due to \tinydtls's processing delays, network buffers might fill up if \aflnet sends too much data in a short time period.
To deal with this, \aflnet checks on every send and receive if all the bytes are sent, and retries if not.

\snapfuzz handles all these issues through its network fuzzing protocol.
(We also note that \tinydtls exercises \snapfuzz's UDP translation capabilities, unlike the other servers which use TCP.)
The end result is that all these delays are eliminated:
\aflnet doesn't need to guess the state of the target anymore, as \snapfuzz explicitly informs \aflnet about the next action of the target.
Similarly, the issue of dropped packets disappears, as \aflnet is always informed when it is the right time to send more data.
Finally, \snapfuzz's UNIX domain sockets eliminate the need for send and receive retries, as full buffer delivery from and to the target is guaranteed by the domain socket protocol.
\tinydtls writes a lot of data to \code{stdout}, so it also benefits from \snapfuzz's ability to skip these system calls.


The one million \tinydtls iterations take on average \tinydtlsAnT under \aflnet, while only \tinydtlsSfT under \snapfuzz, providing a \tinydtlsSfAnSu speedup.

We remind the reader that in \tinydtls we decided to decrease the manually-added time delay between requests from 30ms to 2ms, as we noticed the performance of \aflnet was seriously affected by it. 
Without this change, \aflnet would take significantly longer to complete one million iterations, and the speedup achieved by \snapfuzz would be significantly higher.

\subsection{Performance Breakdown}

\begin{table*}[t]
      \centering
      \caption{Speedup achieved by \snapfuzz compared to \aflnet, when each \snapfuzz component is added one by one.
      Note that the ordering has an impact on the speedup achieved by each component (see text).}
      \captionsetup{justification=centering}
      \label{tbl:breakdown}
      \begin{tabular}{l|r||r|r|r|r|r|}
                      & SF Protocol      & + Affinity       & + No Sleeps       & + No STDIO        & +  Defer         & + In-Mem FS       \\ \hline
            \dcmqrscp & \dcmqrscpSfPAnSu & \dcmqrscpSfAAnSu & \dcmqrscpSfSlAnSu & \dcmqrscpSfStAnSu & \dcmqrscpSfDAnSu & \dcmqrscpSfFsAnSu \\ \hline
            \dnsmasq  & \dnsmasqSfPAnSu  & \dnsmasqSfAAnSu  & \dnsmasqSfSlAnSu  & \dnsmasqSfStAnSu  & \dnsmasqSfDAnSu  & \dnsmasqSfFsAnSu  \\ \hline
            \tinydtls & \tinydtlsSfPAnSu & \tinydtlsSfAAnSu & \tinydtlsSfSlAnSu & \tinydtlsSfStAnSu & \tinydtlsSfDAnSu & \tinydtlsSfFsAnSu \\ \hline
            \lightftp & \lightftpSfPAnSu & \lightftpSfAAnSu & \lightftpSfSlAnSu & \lightftpSfStAnSu & \lightftpSfDAnSu & \lightftpSfFsAnSu \\ \hline
            \live     & \liveSfPAnSu     & \liveSfAAnSu     & \liveSfSlAnSu     & \liveSfStAnSu     & \liveSfDAnSu     & \liveSfFsAnSu     \\
      \end{tabular}
\end{table*}

In \S\ref{sec:lightftp}--\S\ref{sec:tinydtls} we discuss which components of \snapfuzz are likely to benefit each application the most.
Those conclusions were reached by investigating the system calls issued by the applications, using the estimates provided by strace about how much each syscall takes in the kernel.
To have a better quantitative understanding of the contribution of each components, we performed an ablation study in which we have run different versions of \snapfuzz for a short number of 10k iterations.
We chose a much smaller number of iterations because running so many experiments with 1M iterations was prohibitive on our computing infrastructure.
This means that our speedups sometimes differ significantly from those achieved by 1M iterations.
However, the main goal of these experiments is to gain additional insights into the impact of different components and their interaction. 

Due to various dependencies among components, we start with a version of \snapfuzz containing only the network fuzzing protocol, and keep adding components one by one.
However, it is essential to understand that the order in which we add components matters, as their effect is often multiplicative.
In particular, this means that the additional impact of components added earlier can be significantly diminished compared to the case where the same component is added later.
We give two examples:
\begin{enumerate}[leftmargin=*]
\item \textbf{\snapfuzz protocol and smart affinity.}
The \snapfuzz protocol is a performant non-blocking protocol that polls the fuzzer and the application for communication.
Under the default restricted CPU affinity of \aflnet, the protocol is under-performing, because the polling model requires independent CPU cores to get the expected performance benefit.
At the same time, the smart CPU affinity component depends on whether the \snapfuzz protocol is enabled or not, as the protocol changes what it is executed on the CPU.
\item \textbf{In-memory filesystem and smart deferred forkserver}.
The smart deferred forkserver performs better when the in-memory filesystem is enabled, because with an in-memory filesystem it can delay the forkserver past filesystem operations. 
On the other hand, the in-memory filesystem also performs better when the smart deferred forkserver is enabled.
This is because the in-memory filesystem has a fixed overhead of loading and storing the files the target is reading in the beginning of its execution.
This initial overhead might degrade performance, especially for short executions.
When the deferred forkserver is enabled, this overhead is bypassed, as these files are loaded only once in memory and consecutive operations will be only in-memory.
\end{enumerate}

One option would be to try all possible orderings.
However, the full number is large (6! = 720) and some orderings are difficult to run due to engineering limitations (\eg the \snapfuzz protocol is deeply embedded into \snapfuzz and disabling it would require a major engineering overhaul).
Nevertheless, we believe the ordering we present here is still useful in providing insights into the impact of each \snapfuzz component.

Table~\ref{tbl:breakdown} shows our results.
We observe that all components have a significant impact on at least one benchmark.
Furthermore, the \snapfuzz protocol, the smart  affinity, and the smart deferred forkserver always lead to gains, while eliminating developer-added delays (\textit{no sleeps}), avoiding stdout/stderr writes (\textit{no stdout}) and the in-memory file system make no difference in some benchmarks.
Removing writes to stdout/stderr is the least impactful component, benefiting only \live. 

The reported numbers are largely consistent with our qualitative observations of  \S\ref{sec:lightftp}--\S\ref{sec:tinydtls}.
For instance, the main benefits of \lightftp come from the \snapfuzz protocol (\lightftpSfPAnSu) which removes synchronization and server termination delays; from smart affinity (\lightftpSfAAnSu), as \lightftp is a multi-threaded application; from removing developer-added delays, which are present in \lightftp (\lightftpSfSlAnSu); from the smart deferred forkserver (\lightftpSfDAnSu), as it has a long initialization phase; and from the in-memory filesystem (\lightftpSfFsAnSu), as it makes heavy use of the filesystem.
While \lightftp has writes to stdout, removing them does not make a noticeable difference. 

The performance numbers for other benchmarks also largely agree with our expectations.
For instance, the in-memory filesystem brings no benefits to \dnsmasq, which is an in-memory database with little filesystem interaction;
but it highly benefits from the smart deferred server (\dnsmasqSfDAnSu), given that it has a long initialization with over 1,200 system calls issued before it is able to accept input.

\subsection{Unique Crashes Found}



\snapfuzz, as expected, was able to replicate all \aflnet discovered crashes.
Through its performance advantage, it also found additional crashes in 3 of the 5 benchmarks.
During 24h fuzzing campaigns, \snapfuzz found 4 bugs in the \dcmqrscp benchmark while \aflnet was not able to find any.
For \dnsmasq, \snapfuzz was able to find 7 crashes while \aflnet found only 1, and for the \live benchmark, \snapfuzz was able to find 4 crashes while \aflnet found 2.
Both tools found 3 bugs in \tinydtls.
Overall, \snapfuzz found 18 unique crashes, 12 more than \aflnet.

The bugs are a variety of heap overflows, stack overflows, use-after-free bugs, and other types of undefined behaviours.
Fortunately, they seem to have been fixed in the latest versions of these applications.
We plan to rerun \snapfuzz on the latest versions. 

\section{Related Work}
\label{sec:related}
\snapfuzz focuses on creating an efficient fuzzing platform for network applications and helps algorithmic research to be built on top of a strong foundation.
We envision that this separation of concerns will help future research to progress faster by alleviating the laborious task of building performant fuzzers for network and other stateful applications.

\snapfuzz builds on top of \aflnet~\cite{aflnet}, and reuses its ability to infer network protocols.
However, \aflnet has various inefficiencies and requires fragile manual delays and clean-up scripts in its fuzzing harnesses.
Our comprehensive evaluation against \aflnet shows how \snapfuzz can address both problems, resulting in impressive speedups in the range of \minSu-\maxSu.

Besides \aflnet, a popular way of fuzzing network applications is via the \textit{de-socketing} functionality of \preeny~\cite{preeny}.
\preeny intercepts networking functions such as \code{connect} and \code{accept} and makes them return sockets that are synchronised with \code{stdin} and \code{stdout}, essentially allowing \afl to continue to fuzz files and redirecting their contents over network sockets, as expected by the network applications being tested.
Synchronisation is done in a hacky way:
\preeny implements a small server thread that is continuously polling \afl's generated input file and then forwards the read data to the appropriate network calls through a UNIX domain socket to the target~\cite{preeny-extra-thread}.
While a direct comparison with \aflnet and \snapfuzz is not easily possible because a meaningful fuzzing campaign requires the network protocol inferred by \aflnet, we expect a rewrite of \aflnet on top of \preeny to be slower than vanilla \aflnet, due to the extra overhead imposed by file-based fuzzing and the additional thread server used by \preeny.

Multifuzz~\cite{multifuzz} presents a more advanced de-socketing library called Desockmulti, which is similar to \preeny, but optimized in various ways, \eg by removing the use of threads and adding the ability to initiate multiple connections to the target.
\multifuzz is specifically designed for publish/subscribe protocols and the evaluation does not include the benchmarks used by \aflnet and us.
For the two benchmarks used, libcoap and Mosquitto, the paper reports throughput increases of 62.6x and 147.6x respectively on top of \aflnet.
We expect \snapfuzz to perform better particularly due to its specialized network fuzzing protocol and its memory file system (\multifuzz uses \code{tmpfs}, see \S\ref{sec:state-reset}) but unfortunately, \multifuzz is not available as open source (only its Desockmulti library is available), so a direct comparison is not possible.

Xu et al.~\cite{os-primitives-fuzzing} propose new operating systems primitives for fuzzing.
This include, for instance, a new \texttt{snapshot} system call, which aims to address the same goal as \snapfuzz with respect to efficiently snapshotting the target.
The main disadvantage of this approach is that it requires kernel support; by contrast, \snapfuzz runs in user mode, using an unmodified OS. 

Most work on testing network protocol implementations has focused on algorithmic rather than platform-level improvements, focusing in particular on inferring network protocol implementations~\cite{aflnet,polyglot,tupni,sgpfuzzer}.
This work is orthogonal to \snapfuzz and could be combined with it, as we have done with \aflnet's protocol inference algorithm.
More broadly, greybox fuzzing is an active area of research~\cite{fuzzing:ieee-sw} with recent work on improving its effectiveness by directing exploration toward interesting program parts~\cite{aflgo,aflfast}, combining it with symbolic execution~\cite{stephens2016driller,munch,savior}, inferring input grammars~\cite{nautilus,superion} or specialising it to various application domains~\cite{jfs,pgfuzz,squirrel,snooze}.

Besides greybox fuzzing, other forms of fuzzing have been used to test network applications, in particular blackbox fuzzing~\cite{pulsar,mlnetfuzzing} and fault injection~\cite{network-emulator,lfi}.
%
%

%

\section{Conclusion}
\label{sec:conclusion}

Fuzzing stateless applications has proven extremely successful, with hundreds of bugs and security vulnerabilities being discovered.
Recently, in-depth fuzzing of stateful applications such as network servers has become feasible, due to algorithmic advances that make it possible to generate inputs that follow the application's network protocol.
Unfortunately, fuzzing such applications requires clean-up scripts and manually-configured time delays that are error-prone, and suffers from low fuzzing throughput.
\snapfuzz addresses both challenges through a robust architecture, which combines a synchronous communication protocol with an in-memory filesystem and the ability to delay the forkserver to the latest safe point, as well as other optimizations.
As a result, \snapfuzz simplifies fuzzing harness construction and improves the fuzzing throughput significantly, between \minSu and \maxSu on a set of popular network applications, allowing it to find additional crashes.

\snapfuzz will be submitted for artifact evaluation and made available to the community as open-source shortly after publication, with the hope that it will help improve the security and reliability of network applications and facilitate further research in this space.


\balance
\bibliographystyle{ACM-Reference-Format}
\bibliography{ms}

\end{document}